# Implementation of a Wake-up Radio Cross-Layer Protocol in OMNeT++ / MiXiM

Jean Lebreton and Nour Murad

University of La Reunion, LE2P

40 Avenue de Soweto, 97410 Saint-Pierre

Email: jean.lebreton@univ-reunion.fr

*Abstract*—This paper presents the DoRa protocol, which is a new cross-layer protocol for handling the double radio of nodes in wake-up radio scenario. The implementation details in OMNET++/MiXiM are also given, with a focus on the implemented MAC layers. The main goal of the DoRa protocol is to reduce energy consumption in wireless sensor network, by taking full advantage of the passive wake-up scheme. The performance of the DoRa protocol is then evaluated and results are compared with B-MAC and IEEE 802.15.4 protocols.

## I. INTRODUCTION

Energy conservation in Wireless Sensor Networks (WSN) is essential since nodes are usually powered by limited batteries. Several approaches have been considered in the literature to reduce the energy consumption of the network [1]. Duty cycling mechanisms are part of the proposed approaches and they can prolong the nodes lifetime consequently. In duty cycling mechanisms, nodes alternate between different states (typically awake and sleep) in order to avoid idle listening, which is the major source of energy depletion in WSN. We propose to reduce even more the duty cycle with a new wake-up radio cross-layer protocol, which means the nodes would wake up on demand by a particular radio message.

A second radio module is plugged into the nodes for the wake-up communications and the main radio module is used for data transmission. The second radio, also known as wake-up radio, must be a low-power device or completely passive. Few MAC protocols were proposed during the last years to incorporate the wake-up radio features, such as RTM [2], On Demand MAC [3] and SCM-MAC [4]. These protocols reduce significantly the energy consumption of the nodes but they do not take full advantage of the wake-up scheme. Our MAC protocol takes into account the fact that communication is done while other nodes are sleeping, enabling the non-use of RTS/CTS mechanisms or acknowledgements.

This paper presents the implementation of a new MAC protocol in OMNeT++ 4.6 with the MiXiM 2.3 framework [5],[6]. The node architecture has been modified to enable the communication with two radio modules and cross-layer information exchanges, as the common node model is not appropriate for our proposed MAC protocol. Besides, the B-MAC [7] and IEEE 802.15.4 MAC protocol [8] are already implemented in MiXiM, which enables the performances comparison with existing protocols.

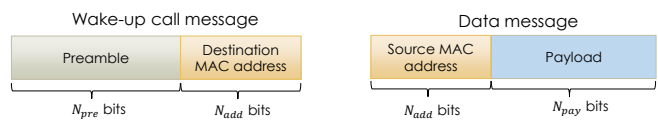

Fig. 1. Frame structure of exchanged messages between the BS and nodes.

## II. DoRa PROTOCOL DESIGN

The DoRa protocol is designed for hierarchical networks composed of a Base Station (BS) and $N$ nodes, in which a one-hop communication is used between the BS and nodes. The BS is in charge of every wake-up procedures and it triggers the nodes wake-up one at a time, which prevents potential communication conflicts if they would wake themselves up. This mechanism is more energy-efficient from nodes perspective since they can directly communicate after waking up. Some hardware modifications are required on the nodes to be compliant with the DoRa protocol. A second low-power micro-controller is used with a passive radio receiver, which can harvest energy from received signals. These two main modules defines our Wake-up Radio (WuR).

All nodes are initially in sleeping mode with their main radio card switched off. The BS sends a wake-up call periodically to each node by including its MAC address in the wake-up call message. The awake node answers by sending the data message directly to the BS. Both frame structures are depicted in figure 1. Before sending a new wake-up call, the BS waits during a predefined timeout $T_{out}$ or until the data reception from the concerned node. Every node is requested with this communication scheme and no internal interference is possible since nodes stay in sleep mode while they do not receive a wake-up call addressed to them. The communication principle is illustrated in figure 2.

This protocol implies that the BS alternates between transmission and reception mode. The reception state is activated once a wake-up call is sent and the transmission states is activated when the wake-up call must be sent. Concerning the node, its main radio has to switch between sleeping mode and transmitting mode, the latter being activated after the wake-up radio has confirmed a wake-up call. Two main criteria must be fulfil for confirming a wake-up call in our scheme. First, the received power of a signal should be strong enough all along the wake-up preamble duration, corresponding to a continuous voltage delivered by a passive wake-up circuit. The second condition is the validation of the MAC address included in the wake-up call message.

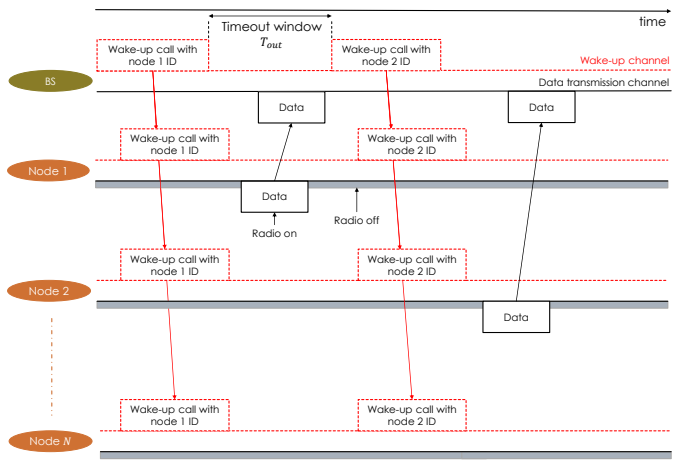

Fig. 2. Communication principle with the DoRa protocol.

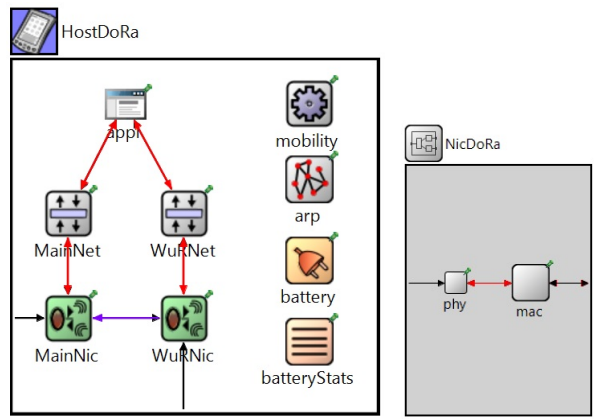

Fig. 3. Structure of *HostDoRa*.

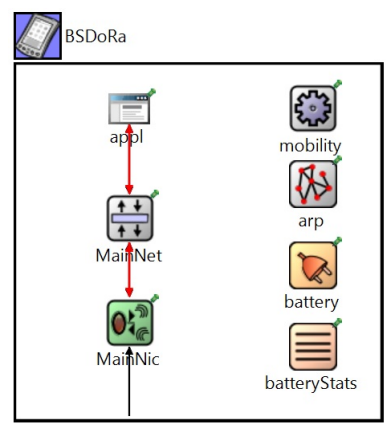

Fig. 5. Structure of *BSDoRa*.

## III. IMPLEMENTATION

This section provides the implementation details of the DoRa protocol in OMNeT++/MiXiM. The node and BS modules are different in our simulation since they have distinct architectures and features. The main contribution lies in the MAC layers used by each module. After describing the main implementation, the finite state machine of each MAC layer is presented.

### A. Node module

The node module is called *HostDoRa* and its overall structure is shown in figure 3. The major implementation is done in both MAC layers, namely *Mac_Main* for handling the communication with the main radio and *DoRaMacLayer* for handling the communication associated with the wake-up process.

- The *DoRaMacLayer* handles wake-up calls sent by the BS. The state machine of the micro-controller, which is in charge of the wake-up procedure, is implemented in this layer. After processing and validating the destination address in a wake-up call, *DoRaMacLayer* sends a control message to both the other MAC layer (*Mac_Main*) and application layer for sending data to the BS.

- The *Mac_Main* handles the state machine of the main radio module. After receiving a control message from *DoRaMacLayer*, it goes to *idle* state and waits for the message from the application layer. The radio is then switched to *Tx* state when the data is received. After transmitting the data through the medium, a control message is sent to *DoRaMacLayer* and the node goes back to *sleep* state.

- A new gate is defined between the MAC layers, enabling a cross-layer communication for control messages. This gate is defined in a new basic module called *BaseLayerCross*. This module is an extension of the *BaseLayer* module, which is the basic layer for every module. There are currently 2 lower gates and 2 upper gates, for data or control messages with the lower or upper layers. We added a side gate for sending control message to an other layer, with the associated function for sending message through the gate. The control messages could be passed through the network and application layers, but a direct communication between the two MAC layers is preferred for saving time in the wake-up procedures. Taking less time for waking up the nodes or going to sleep mode also reduces the energy consumption, which is the main motivation of this work.

- Some modifications were done in the physical layer of MiXiM, especially in the decider. We derived the *Decider802154Narrow* for creating the *DeciderWakeUpRadio* in which a signal is processed if the signal duration is long enough, and with a sufficient energy level all along the signal duration. This type of signal corresponds in fact to our wake-up preamble, which is a distinctive feature from data messages.

### B. Base Station module

The BS module is called *BSDoRa* and its structure is given in figure 5. Only one radio module radio is used for both the wake-up call and the data reception, since they are delivered in the same channel. A MAC layer called *BSMAC* have been implemented for handling the switching between wake-up call transmission and data reception, according to the DoRa protocol. The *BSMAC* is also in charge of incrementing the destination address included in the wake-up call message in order to request data from every nodes.

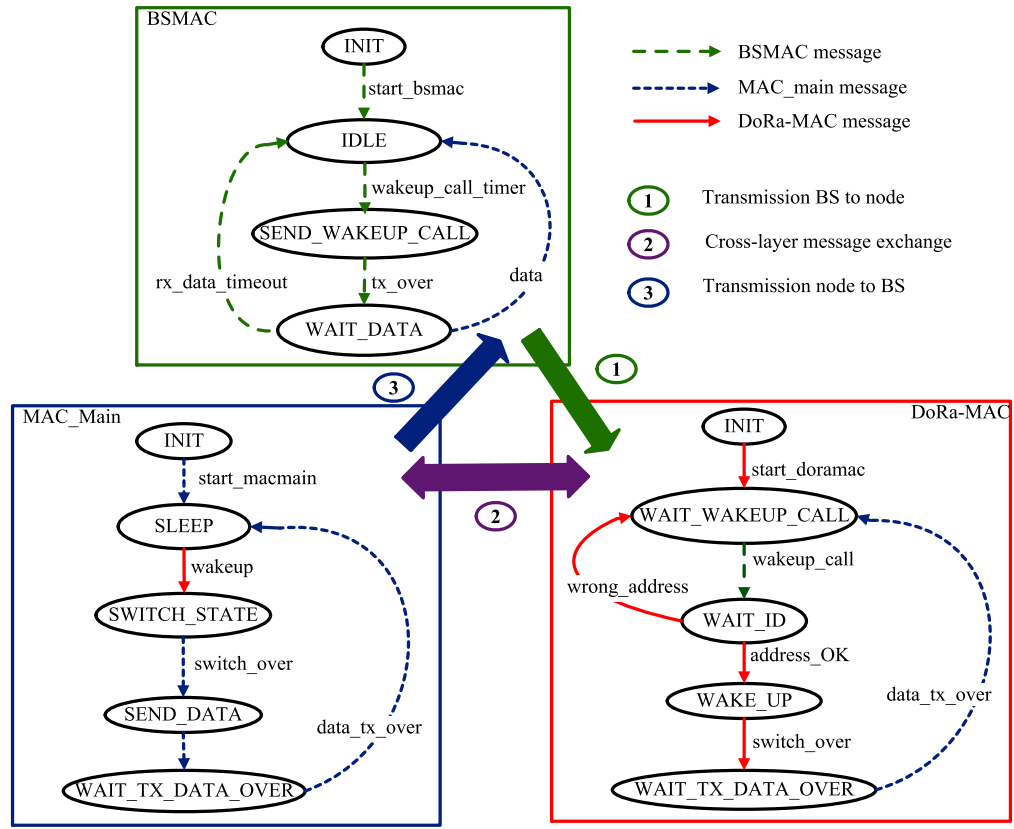

Fig. 4. Finite state machine of each MAC layer of DoRa protocol.

*C. Finite state machines*

The finite state machine of each MAC layer is given in figure 4 and shows the required interactions. Within the node, the interactions between the two MAC layers is the reason of our cross-layer approach. All MAC modules are initialized at the same time and they switch to their first state, waiting for an event to continue. Upon the end of the wake-up call timer, *BSMAC* sends a wake-up call message in the related state. After transmission, the module *BSMAC* waits for data coming from the *MAC_main* module.

The *DoRa-MAC* module switches its state for decoding the MAC address when it receives the wake-up call message. If the destination address included in the message corresponds to its own MAC address, then *DoRa-MAC* switches to *WAKE_UP* state where it sends the control message called *wakeup* to the *MAC_main* module. The last state consists of waiting for the completion of data transmission from the *MAC_main* module.

The *MAC_main* module leaves the *SLEEP* state when it receives the *wakeup* control message, for switching its radio to transmitting mode. After sending data through the channel, the module goes back to *SLEEP* state. At this stage, all modules are back to their first states, since a complete cycle of data request and data transmission is done.

## IV. Performance evaluation

The DoRa protocol is compared with the B-MAC and IEEE 802.15.4 protocols through the same simulation scenario. Even if B-MAC and IEEE 802.15.4 are chosen due to their existing implementation in MiXiM, it gives an interesting comparison of our on-demand wake-up protocol with a duty-cycle approach (B-MAC) and an always on radio approach (IEEE 802.15.4).

A network comprised of a BS and 5 nodes is defined and every node sends data periodically to the BS, with a direct transmission. The inter packet arrival time is the time between two consecutive packets reception and it characterizes the traffic rate. The variation of the inter packet arrival time enables the performance evaluation of the network for each protocol. The other simulations parameters are set with a constant value and the main parameters are given in table I.

A number of 8 simulations are executed by incrementing the inter packet arrival time with each protocol. Even if the performance can be evaluated through several statistics, we

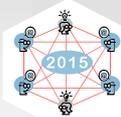

TABLE I
SIMULATION PARAMETERS

| Protocol | Parameter | Value |
|---|---|---|
| Common | Surface area | 20 m × 20 m |
| | Number of nodes | 5 |
| | Packet payload | 100 bytes |
| | Queue buffer length | 10 packets |
| | Supply voltage | 3.3 V |
| | Battery capacity | 2900 mAh |
| | Bit rate | 250 kbps |
| | Sleep current | 900 nA |
| | Rx current | 16.6 mA |
| | Tx current | 21.2 mA |
| IEEE 802.15.4 | Backoff number | 3-5 |
| | CCA detection time | 128 $\mu s$ |
| B-MAC | Slot duration | 1 s |
| | Check interval | 10 ms |
| DoRa-MAC | Sleep current (WuR) | 200 nA |
| | Rx current (WuR) | 1.1 $\mu A$ |
| | Active current (WuR) | 9 mA |
| | Bit rate (WuR) | 32 kbps |
| | Timeout window | 10 ms |

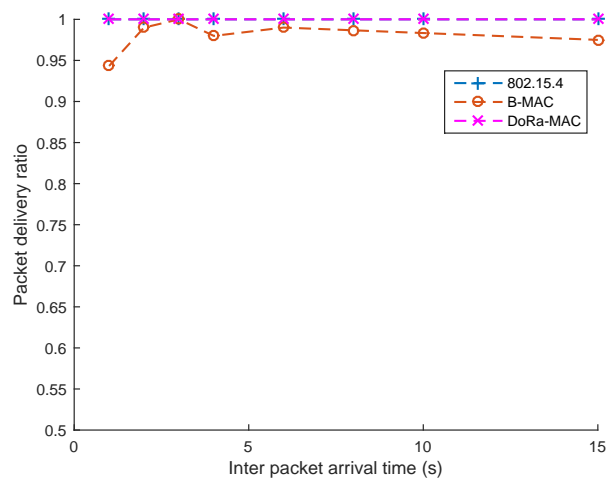

Fig. 7. Packet delivery ratio versus the inter packet arrival time for each protocol.

The results presented in figure 6 can also be used to give the amount of energy saving for B-MAC and DoRa-MAC compared to IEEE 802.15.4. Considering an inter packet arrival time of 15 seconds, B-MAC saves up 95% of energy while DoRa-MAC saves up 99.98% of energy, both compared to 802.15.4 MAC protocol.

The figure 7 gives the packet delivery ratio at the BS, defined as the number of packet received at the BS divided by the number of packets generated by the nodes. The results are consistent for the IEEE 802.15.4 and DoRa protocol since the packet delivery ratio is always equal to 1, while it is comprised between 0.94 and 1 for the B-MAC protocol. This graph shows the reliability of our DoRa-protocol since all the packets are received.

## V. CONCLUSION

The main motivation of the DoRa protocol is to reduce energy consumption in WSN, by providing a cross-layer protocol for passive wake-up radio architecture. The DoRa protocol is presented and its implementation is described in this paper. Simulations were carried out in order to evaluate the performance of DoRa protocol in comparison with B-MAC and IEEE 802.15.4 protocols. The results clearly show the benefit of using DoRa compared to both B-MAC and IEEE 802.15.4 since it reduces significantly the average energy consumption. The results also show the huge benefits of using a passive wake-up radio architecture compared to an always on radio or duty-cycle approach.

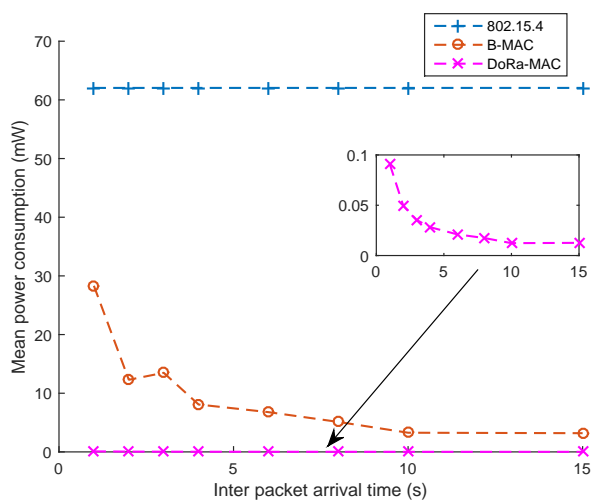

Fig. 6. Mean power consumption versus the inter packet arrival time for each protocol.

focus our study in the mean power consumption of the nodes and packet delivery ratio at the BS.

The mean power consumption is defined as the ratio between the total amount of consumed energy and the total operation time. This mean average power consumption is averaged for every node and its evolution is depicted in figure 6. As no duty-cycle is defined in the IEEE 802.15.4 protocol, the mean power consumption remains constant while varying the inter packet arrival time. The power consumption is relatively high because of continuous idle listening. B-MAC give better results since it takes advantage of duty-cycle for saving noticeable energy compared to IEEE 802.15.4. As depicted in the figure 6, the DoRa protocol outperform the others since the mean power consumption is reduced significantly. This graph also shows that the DoRa is even more suitable for application with low traffic rate, even if it still performs well with higher traffic rate.

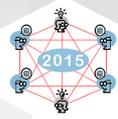